\documentclass[preprintnumbers,11pt,prd,showpacs]{revtex4}
\usepackage{graphicx}
\usepackage{amssymb}
\usepackage{epstopdf}
\usepackage{amsmath}
\begin{document}

\title{The Thomas-Fermi Quark Model: Non-Relativistic Aspects\footnote{Partially supported by a grant from the Baylor University Research Committee.}}

\author{Quan Liu\footnote{quan\_liu@baylor.edu} and Walter Wilcox\footnote{Corresponding author: walter\_wilcox@baylor.edu}}

\affiliation{Department of Physics, Baylor University, Waco, TX 76798-7316}

\begin{abstract}
The first numerical investigation of non-relativistic aspects of the Thomas-Fermi (TF) statistical multi-quark model is given. We begin with a review of the traditional TF model without an explicit spin interaction and find that the spin splittings are too small in this approach. An explicit spin interaction is then introduced which entails the definition of a generalized spin \lq\lq flavor". We investigate baryonic states in this approach which can be described with two inequivalent wave functions; such states can however apply to multiple degenerate flavors. We find that the model requires a spatial separation of quark flavors, even if completely degenerate. Although the TF model is designed to investigate the possibility of many-quark states, we find surprisingly that it may be used to fit the low energy spectrum of almost all ground state octet and decuplet baryons. The charge radii of such states are determined and compared with lattice calculations and other models. The low energy fit obtained allows us to extrapolate to the six-quark doubly strange {\it H}-dibaryon state, flavor symmetric strange states of higher quark content and possible six quark nucleon-nucleon resonances. The emphasis here is on the {\it systematics} revealed in this approach. We view our model as a versatile and convenient tool for quickly assessing the characteristics of new, possibly bound, particle states of higher quark number content.
\end{abstract}


\maketitle
\newpage

\section{Introduction}\label{Sone}

The aim of the Thomas-Fermi (\lq\lq TF") statistical multi-quark model, introduced in Ref.\cite{wilcox}, is to explore many-quark baryon states including strange matter\cite{witten,farhi}. It is a useful new tool for the quark physicist to quickly assess the possibility of bound and resonance states in preparation for much more detailed and expensive lattice QCD calculations. The TF semi-classical model combines Coulombic quark interactions with a bag model type spherical confinement assumption. The model was seen to be versatile in that a number of widely varying physics scenarios, including non-relativistic, extreme relativistic, massive gluon, or color-flavor locking, can be addressed. The development in Ref.\cite{wilcox}, however, was limited to an investigation of a single quark wave function representing an equal number of $N_f$ mass-degenerate flavors. We would like to build upon the previous results to show how wave functions for unequal numbers of degenerate flavors or non-degenerate masses can be determined through the coupled differential equations. Although the model is designed to be most reliable for many-quark states, we find surprisingly that it may be used to fit the low energy spectrum of baryons. Note that this model is designed only to look for true many-quark bound states and resonances rather than bound nucleonic states like the deuteron. In this sense, the model complements calculations of loosely bound states as in Ref.\cite{meissner}. The traditional TF model has no explicit spin interactions since the spin states are treated as degenerate. In order to produce a realistic spectrum, the model was extended to include an explicit spin splitting term. We introduce spin as a generalized \lq\lq flavor" in our non-relativistic model. This change keeps the quark differential equations unmodified and simply induces an overall scale change in the energy expressions. The non-relativistic nature of the model is a shortcoming, but we view our work as the appropriate initial step toward developing a more realistic relativistic version of the model. However, we note that the quark mass parameters dominate the overall baryon masses and that the overall fits are reasonably good, so the model is self-consistent.

The TF quark model is not meant to compete with more fundamental lattice QCD investigations but to explore phenomenological regions where lattice QCD can not yet go. Lattice calculations are limited by volume considerations for many-quark systems and such applications are computationally expensive. This model attempts to \lq\lq set the stage" for such lattice calculations by examining possible scenarios where interesting existence and structural questions can be addressed. Our goal is not to produce final answers but to uncover physical systematics associated with many-quark states. 

In the next section we will describe some of the changes in the formalism necessary to introduce quark spin as a generalized flavor. We will then present the differential equations and the generalized expressions for the quark energies in Section \ref{Sthree} for states with two inequivalent TF wave functions. More of the formalism for the consideration of spin will then be presented in Section \ref{Sfour}. We will describe the computer program that solves the TF wave function differential equations and present our fit to the baryon spectrum in Section \ref{Sfive}. The baryonic charge radii obtained are compared with lattice and other model calculations. One initial interesting finding is that the flavors in a baryon are naturally separated, even if degenerate in mass. As an application, we will attempt such a search in Section \ref{SSone} for the hypothesized {\it H}-dibaryon state\cite{jaffe}, apply the formalism to high multi-quark strange states in Section\ref{SSthree}, and make connection to nucleon-nucleon resonances in Section\ref{SStwo}.

\section{Mathematical Preliminaries}\label{Stwo}

Let us recall how the TF quark model is formulated. The original flavor and color number, $n^I_i(p_{F})$, and non-interacting quark kinetic energy, ${\cal E}^I_i(p_{F})$, densities
were
\begin{equation}
n^I_i(p_{F})=2\int^{p_F}  \frac{d^3p^I_i}{(2\pi \hbar)^3}=\frac{((p_F)^I_i)^3}{3\pi^2\hbar^3}\label{n},
\end{equation}
and
\begin{equation}
{\cal E}^I_i(p_{F})=2\int^{p_F}  \frac{d^3p^I_i}{(2\pi \hbar)^3}\frac{(p^I_i)^2}{2m}=\frac{(3\pi^2\hbar^3n^I_i(p_F))^{5/3}}{10\pi^2\hbar^3m},\label{E}
\end{equation}
where $(p_F)^I_i$ is the Fermi momenta, the \lq\lq $I$" superscript stands for flavor and the \lq\lq $i$" subscript stands for color. Notice the explicit degenerate spin factors of $2$. Such an assumption of degeneracy is excellent for atomic physics, marginal for nuclear physics, but unrealistic for hadronic physics. The addition of a spin term in the TF Hamiltonian requires the introduction of non-equal spin densities. Therefore, we will define new quantities distinguished by caligraphic superscripts,
\begin{equation}
n^{\cal I}_i(p_{F})\equiv \int^{p_F}  \frac{d^3p^{\cal I}_i}{(2\pi \hbar)^3}=\frac{((p_F)^{\cal I}_i)^3}{6\pi^2\hbar^3}\label{n2},
\end{equation}
\begin{equation}
{\cal E}^{\cal I}_i(p_{F})\equiv \int^{p_F}  \frac{d^3p^{\cal I}_i}{(2\pi \hbar)^3}\frac{(p^{\cal I}_i)^2}{2m_{\cal I}}=\frac{(6\pi^2\hbar^3n^{\cal I}_i(p_{F}))^{5/3}}{20\pi^2\hbar^3m_{\cal I}},\label{E2}
\end{equation}
where the spin specification is now included in an augmented flavor-spin index, ${\cal I}$. (For notational simplicity, the arguments of $n^{\cal I}_i$ and ${\cal E}^{\cal I}_i$ have been simplified from $((p_F)^{\cal I}_i)$ to $p_F$.) We are allowing for different masses, $m_{\cal I}$, and different Fermi momentums for each flavor, $p_F^{\cal I}$. After variation of the densities in the energy functional, Eq.(21) of Ref.\cite{wilcox} still holds:
\begin{equation}
\frac{(p_F^{\cal I})^2} {2m_{\cal I}}=-\lambda^{\cal I} +
\frac{3\times\frac{4}{3}g^2}{(3A-1)}\left(\frac{N_{\cal I}-1}{N_{\cal I}} \int^{r_{max}}\!\!d^3r' 
\frac{n^{\cal I}(r')}{|{\vec r}-{\vec r}\,'|}+\sum_{{\cal J}\ne {\cal I}}\int^{r_{max}}\!\!d^3r' \frac{n^{\cal J}(r')}{|{\vec r}-{\vec r}\,'|} \right). \label{Equation}
\end{equation}
We have assumed as before that the color densities are the same: $n^{\cal I}\equiv n^{\cal I}_1=n^{\cal I}_2=n^{\cal I}_3$. The momentum density, $(p_F^{\cal I})^2 /2m_{\cal I}$, is defined in terms of particle density from Eq.(\ref{n2}). However, the previous relationship between the TF spatial wave function and the density,
\begin{equation}
f^{I}(r) \equiv   \frac{ra}{2\times\frac{4}{3}\alpha_s}(3\pi^2 n^{I}(r))^{2/3}\label{FinN1}.
\end{equation}
is now
\begin{equation}
f^{\cal I}(r) \equiv   \frac{ra}{2\times\frac{4}{3}\alpha_s}(6\pi^2 n^{\cal I}(r))^{2/3}\label{FinN2}.
\end{equation}
The fundamental scale $a$ is 
\begin{equation}
a\equiv \frac{\hbar}{m_1c},
\end{equation}
where $m_1$ is the lightest quark mass.  Rather than change the TF differential equations it turns out that all one needs do to accommodate the new approach is to make a change in the overall spatial scale factor. The previous definition
\begin{equation}
r = Rx, \quad\quad R \equiv \left(\frac{a}{2\times\frac{4}{3}\alpha_s}\right)\left[\frac{3\pi A}{4} \right]^{2/3},\label{peq1}
\end{equation}
is now replaced by
\begin{equation}
r = {\cal R}x, \quad\quad{\cal R} \equiv \left(\frac{a}{2\times\frac{4}{3}\alpha_s}\right)\left[\frac{3\pi A}{2} \right]^{2/3}.\label{peq2}
\end{equation}
equivalent to a simple change in the underlying scale factor:
\begin{equation}
a\longrightarrow 2^{2/3}a.\label{scalechange}
\end{equation}
The form of the differential TF equations in Ref.\cite{wilcox} and the wave function normalization,
\begin{equation}
\int_0^{x_{max}} dx \sqrt{x}(f^{\cal I}(x))^{3/2}=\frac{N_{\cal I}}{3A}.\label{norm}
\end{equation}
are unchanged from before.

\section{Generalized Model for Two Wave Functions}\label{Sthree}
In Ref.\cite{wilcox}, a non-relativistic model with equal numbers of $N_f$ degenerate flavors was considered in a case study. However, in order to do realistic spectrum calculations, we have to generalize to unequal numbers of flavors. In this section, we explain how to solve for systems with two inequivalent TF wave functions. This can arise either from flavors with different masses or systems with two equal masses but different numbers of particles. We will see this is equivalent to the introduction of two quark flavor-degeneracy factors, $(N_f)_1\equiv g_1$ and $(N_f)_2\equiv g_2$. 

The differential form of the TF quark equations for this case, which involves two distinct wave functions, is
\begin{eqnarray}
\alpha_{\cal I}\frac{d^2f^{\cal I}(x)}{dx^2}=-\frac{A}{A-\frac{1}{3}}\frac{1}{\sqrt{x}}\left(\frac{N_{\cal I}-1}{N_{\cal I}}\left(f^{\cal I}(x)\right)^\frac{3}{2}+\sum_{{\cal J}\ne {\cal I}}\left(f^{\cal I}(x)\right)^\frac{3}{2}\right).\label{generalform}
\end{eqnarray}
Here $\alpha_{\cal I}$ is the mass ratio of the lightest mass quark, $m_1$, to quark flavor $\cal I$,
\begin{eqnarray}
\alpha_{\cal I}=\frac{m_1}{m_{\cal I}}.\label{alpha}
\end{eqnarray}
and $N_{\cal I}$ is the number of quarks with generalized flavor ${\cal I}$. (Note we are using subscripts on $N_{\cal I}$, $\alpha_{\cal I}$ and $m_{\cal I}$ so that they will not be mistaken for powers. In Ref.\cite{wilcox} all these quantities were denoted with superscripts.) For $g_{1}$ flavors with equal number $ N_1$, considered the state with the larger radius, $x_1\ge x_2$, and $g_{2}$  flavors with equal numbers $ N_2$, we have
\begin{equation}
g_{1}N_1+g_{2}N_2=3A, \label{add}
\end{equation}
for quark quantum numbers, and the generalized two wave function equations are:
\begin{eqnarray}
\alpha_{1}\frac{d^2f^1(x)}{dx^2} =-\frac{A}{A-\frac{1}{3}}\frac{1}{\sqrt{x}}\left[\left(\frac{N_1-1}{N_1}+g_{1}-1\right)\left(f^1(x)\right)^{3/2}+g_{2}\left(f^2(x)\right)^{3/2}\right],\label{one}\\
\alpha_{2}\frac{d^2f^2(x)}{dx^2} =-\frac{A}{A-\frac{1}{3}}\frac{1}{\sqrt{x}}\left[\left(\frac{N_2-1}{N_2}+g_{2}-1\right)\left(f^2(x)\right)^{3/2}+g_{1}\left(f^1(x)\right)^{3/2}\right].\label{two}
\end{eqnarray}
In the spirit of the TF atomic model, we assume there is only one truly independent wave function. In this two wave function case, we assume a linear relation between $f^1(x)$ and $f^2(x)$:  $f^1(x)=kf^2(x)$. The equations (\ref{one}) and (\ref{two}) now become:
\begin{eqnarray}
\alpha_{1}k\frac{d^2f^2(x)}{dx^2} =-\frac{A}{A-\frac{1}{3}}\frac{1}{\sqrt{x}}\left[\left(\frac{N_1-1}{N_1}+g_{1}-1\right)k^{3/2}+g_{2}\right]\left(f^2(x)\right)^{3/2},\label{18}\\
\alpha_{2}\frac{d^2f^2(x)}{dx^2} =-\frac{A}{A-\frac{1}{3}}\frac{1}{\sqrt{x}}\left[\left(\frac{N_2-1}{N_2}+g_{2}-1\right)+g_{1}k^{3/2}\right]\left(f^2(x)\right)^{3/2}.\label{19}
\end{eqnarray}
To make these two equation consistent with each other implies that
\begin{eqnarray}
\Rightarrow\alpha_{1}g_{1}k^{5/2}-\alpha_{2}\left(\frac{N_1-1}{N_1}+g_{1}-1\right)k^{3/2}+\alpha_{1}\left(\frac{N_2-1}{N_2}+g_{2}-1\right)k-\alpha_{2}g_{2}=0.\label{consistent}
\end{eqnarray}
We solve this condition numerically for the $k$ value for each particle parameter set.

Remember we have two normalization conditions:
\begin{equation}
\left\{ \begin{array}{ll}
               \int_{0}^{x_{2}}\sqrt{x}\left(f^2(x)\right)^{3/2}dx=(N_2/3A),      \\
               \int_{0}^{x_{1}}\sqrt{x}\left(f^1(x)\right)^{3/2}dx=(N_1/3A).
              \end{array}
\right.\label{five}
\end{equation}
With Eq.(\ref{five}), we can express our normalization conditions in the form of boundary conditions:
\begin{equation}
\left\{ \begin{array}{ll}
             \left(x\frac{df^2(x)}{dx}-f^2(x)\right)|_{x_{2}}=-\frac{A}{A-\frac{1}{3}}\left[g_{1}k^{3/2}+\left(\frac{N_2-1}{N_2}+g_{2}-1\right)\right]\frac{N_2}{3A\alpha_{2}},     \\
             \left(x\frac{df^1(x)}{dx}-f^1(x)\right)|_{x_{1}}=-\frac{1}{\alpha_{1}}.
              \end{array}
\right.
\end{equation}
The $f^1(x)=k f^2(x)$ condition must be consistent with the normalization conditions. We will see that this in general implies that the $f_2$ TF wave function has an internal discontinuity or discontinuities. The position or positions of these are not determined by the TF quark differential equations, but by energy minimization. Because of the attractive nature of the system, it is a natural assumption that the wave functions are continuously connected to the origin. That is, we assume no voids in the particle interiors. Although this is a natural assumption, we have not proven these configurations always have the lowest energies.

With these modified boundary conditions, we can derive the expression for kinetic, $T$, and potential, $U$, energies. For the kinetic energy one starts with
\begin{eqnarray}
T=\sum_{{\cal I},i}{\cal E}^{\cal I}_i(p_{F}^{\cal I})=3\sum_{{\cal I}}{\cal E}^{\cal I}(p_{F}^{\cal I}),
\end{eqnarray}
where ${\cal E}^{\cal I} \equiv {\cal E}^{\cal I}_1={\cal E}^{\cal I}_2={\cal E}^{\cal I}_3$. Using Eqs.(\ref{n2}), (\ref{E2}) and (\ref{FinN1}) then gives
\begin{eqnarray}
T=\sum_{{\cal I}}\frac{12}{5\pi}\left(\frac{3\pi A}{2}\right)^{1/3}\frac{\frac{4}{3}g^{2} \times \frac{4}{3}\alpha_{s}}{a}\alpha_{\cal I} \int_{0}^{x_{{\cal I}}}\frac{\left(f^{\cal I}(x)\right)^{5/2}}{\sqrt{x}}dx.
\end{eqnarray}
In our case with $g_1$ flavors with $N_1$ particles and $g_2$ flavors with $N_2$ particles we have
\begin{eqnarray}
T=\frac{12}{5\pi}\left(\frac{3\pi A}{2}\right)^{1/3}\frac{\frac{4}{3}g^{2} \times \frac{4}{3}\alpha_{s}}{a}\left[ g_{1}\alpha_1\int_{0}^{x_{1}}\frac{\left(f^1(x)\right)^{5/2}}{\sqrt{x}}dx+g_{2}\alpha_2\int_{0}^{x_{2}}\frac{\left(f^2(x)\right)^{5/2}}{\sqrt{x}}dx \right ].
\end{eqnarray}
Using the wave function differential equations, consistency condition for $k$ and boundary conditions allows one to relate the integrals to wave function values and derivatives on the discontinuous surfaces:
\begin{eqnarray}
\lefteqn{T=\frac{12}{5\pi}\left(\frac{3\pi A}{2}\right)^{1/3}\frac{\frac{4}{3}g^{2} \times \frac{4}{3}\alpha_{s}}{a}\left[-\frac{5}{7}g_{2}\frac{df^2(x)}{dx}|_{x_{2}}\frac{N_2}{3A}\left(\alpha_2-\frac{kg_{1}\alpha_1}{\frac{N_1-1}{N_1}+g_{1}-1}\right)\right.}\nonumber  \\
& &\left.-\frac{5}{7}\frac{\frac{A-\frac{1}{3}}{A}g_{1}\alpha_1}{\frac{N_1-1}{N_1}+g_{1}-1}\frac{df^1(x)}{dx}|_{x_{1}}+\frac{4}{7}g_{2}\alpha_2\left(f^2(x_{2})\right)^{5/2}\sqrt{x_{2}}+\frac{4}{7}g_{1}\alpha_1\left(f^1(x_{1})\right)^{5/2}\sqrt{x_{1}}\right].\label{fullkinetic}
\end{eqnarray}
Likewise for the potential energy we begin with
\begin{eqnarray}
U= -\frac{ 9\times\frac{4}{3}g^2}{2(3A-1)}\left[\sum_{\cal I} \frac{N_{\cal I}-1}{N_{\cal I}} \int^{r_{\cal I}}\int^{r_{\cal I}}d^3r\,d^3r'\frac{n^{\cal I}(r)n^{\cal I}(r')}{|{\vec r}-{\vec r}\,'|}\right. \nonumber \\
+\left.\sum_{{\cal I}\ne{\cal J}}\int^{r_{\cal I}}\int^{r_{\cal J}}d^3r\,d^3r'\frac{n^{\cal I}(r)n^{\cal J}(r')}{|{\vec r}-{\vec r}\,'|}\right].
\end{eqnarray}
Reducing this to the TF wave functions, $f^{\cal I}(x)$, gives
\begin{eqnarray}
\lefteqn{U=-\frac{2}{\pi}\frac{A}{A-\frac{1}{3}}\left(\frac{3\pi A}{2}\right)^{1/3}\frac{\frac{4}{3}g^{2} \times \frac{4}{3}\alpha_{s}}{a} } \nonumber \\
&&\times\left[ \sum_{\cal I}\frac{N_{\cal I}-1}{N_{\cal I}}\left[\int_{0}^{x_I}dx\frac{\left(f^{\cal I}(x)\right)^{3/2}}{\sqrt{x}}\int_{0}^{x}dx'\sqrt{x'}(f^{\cal I}(x'))^{3/2}+\int_{0}^{x_{\cal I}}dx\sqrt{x}(f^{\cal I})^{3/2}(x)\int_{x}^{x_{\cal I}}dx'\frac{\left(f^{\cal I}(x')\right)^{3/2}}{\sqrt{x'}}\right]\right. \nonumber \\
& &+\left. \sum_{\cal I \neq \cal J}\left[\int_{0}^{x_{\cal I}}dx\frac{\left(f^{\cal I}(x)\right)^{3/2}}{\sqrt{x}}\int_{0}^{x}dx'\sqrt{x'}(f^{\cal J}(x'))^{3/2}+\int_{0}^{x_{\cal I}}dx\sqrt{x}(f^{\cal I}(x))^{3/2}\int_{x}^{x_{\cal J}}dx'\frac{\left(f^{\cal J}(x')\right)^{3/2}}{\sqrt{x'}}\right]\right].
\end{eqnarray}
In the ($g_1, N_1$), ($g_2, N_2$) case this gives
\begin{equation}
U=-\frac{2}{\pi}\frac{A}{A-\frac{1}{3}}\left(\frac{3\pi A}{2}\right)^{1/3}\frac{\frac{4}{3}g^{2} \times \frac{4}{3}\alpha_{s}}{a}
\left[ g_1(\frac{N_1-1}{N_1}+g_1-1)K_1 +g_2(\frac{N_2-1}{N_2}+g_2-1)K_2 +2g_1g_2 K_{12}\right],
\end{equation}
where
\begin{equation}
K_1 \equiv \int_{0}^{x_1}dx\frac{\left(f^1(x)\right)^{3/2}}{\sqrt{x}}\int_{0}^{x}dx'\sqrt{x'}(f^1(x'))^{3/2}+\int_{0}^{x_1}dx\sqrt{x}(f^1(x))^{3/2}\int_{x}^{x_1}dx'\frac{\left(f^1(x')\right)^{3/2}}{\sqrt{x'}},
\end{equation}
\begin{equation}
K_2 \equiv  \int_{0}^{x_2}dx\frac{\left(f^2(x)\right)^{3/2}}{\sqrt{x}}\int_{0}^{x}dx'\sqrt{x'}(f^2(x'))^{3/2}+\int_{0}^{x_2}dx\sqrt{x}(f^2(x))^{3/2}\int_{x}^{x_2}dx'\frac{\left(f^2(x')\right)^{3/2}}{\sqrt{x'}},
\end{equation}
and
\begin{equation}
K_{12} \equiv  \int_{0}^{x_1}dx\frac{\left(f^1(x)\right)^{3/2}}{\sqrt{x}}\int_{0}^{x}dx'\sqrt{x'}(f^2(x'))^{3/2}+\int_{0}^{x_1}dx\sqrt{x}(f^1(x))^{3/2}\int_{x}^{x_2}dx'\frac{\left(f^2(x')\right)^{3/2}}{\sqrt{x'}}.
\end{equation}
(A nontrivial point is that switching $1 \rightleftarrows 2$ above does not change the $K_{12}$ integral.) Again, doing the integrals as above gives the amazingly compact result,
\begin{eqnarray}
\lefteqn{U=-\frac{2}{\pi}\left(\frac{3\pi A}{2}\right)^{1/3}\frac{\frac{4}{3}g^{2} \times \frac{4}{3}\alpha_{s}}{a}  \times \left[\frac{N_{2}}{3A}\frac{df^{2}(x)}{dx}|_{x_{2}}\left(-\frac{12}{7}\alpha_{2}g_{2}+\frac{12}{7}\frac{g_{1}g_{2}k\alpha_{1}}{\frac{N_{1}-1}{N_{1}}+g_{1}-1}\right)\right.} \nonumber \\
& &+\left.\frac{df^{1}(x)}{dx}|_{x_{1}} \left(-\frac{g_{1}g_{2}\alpha_{1}}{\frac{N_{1}-1}{N_{1}}+g_{1}-1}\frac{N_{2}}{3A}-k^{3/2}g_{1}\alpha_{1}\frac{N_{2}}{3A}-\frac{5}{7}\frac{\frac{A-1/3}{A}\alpha_{1}g_{1}}{\frac{N_{1}-1}{N_{1}}+g_{1}-1}\right) \right.\nonumber \\
& &+\left.\frac{\frac{A-1/3}{A}\alpha_{1}g_{1}}{\frac{N_{1}-1}{N_{1}}+g_{1}-1}-\frac{N_{2}}{3A}\alpha_{1}g_{1}k^{3/2}-\frac{g_{1}g_{2}\alpha_{1}}{\frac{N_{1}-1}{N_{1}}+g_{1}-1}\frac{N_{2}}{3A}\right. \nonumber \\
& &+\left.\frac{4}{7}g_{2}\alpha_{2}\left(f^{2}(x_{2})\right)^{5/2}\sqrt{x_{2}}+\frac{4}{7}g_{1}\alpha_{1}\left(f^{1}(x_{1})\right)^{5/2}\sqrt{x_{1}}\right],\label{fullpotential}
\end{eqnarray}
which likewise reduces $U$ to values and derivatives of the TF wave functions at the surfaces.

Of course, one is not limited to two different wave functions in this model, and the above equations can be generalized. However, we will see that such a generalization is not necessary for the low energy baryon spectrum for mass degenerate light quarks.

\section{More on Spin}\label{Sfour}

As outlined above, the traditional TF model simply assumes a spin degeneracy factor of 2 in Eqs.(\ref{n}) and (\ref{E}). We initially attempted to develop such a model based upon Ref.\cite{wilcox}. In this traditional approach there is an intrinsic splitting between states like the nucleon and $\Delta^{++}$ (or $\Delta^{-}$) already just from the different TF flavor wave functions, exclusive of spin, with the correct ordering of states. That is, there is a degeneracy splitting from the different quark flavor content, $uud$ versus $uuu$, for example. However, the nucleon-delta splitting turns out to be much smaller than the actual splitting for reasonable parameter values. In our attempts to fit the low energy baryon spectrum, we found that our $\alpha_s$ and $m_1$ parameters (the latter sets the overall scale) were driven to unrealistic values to try to account for such splittings. The model was incapable of producing a realistic low energy spectrum. It is necessary to add an explicit spin interaction term.

This introduces a problem of course because products of the spin \lq\lq up" and \lq\lq down" (usually taken along the z-axis) in the semi-classical TF model do not combine to form \lq\lq good" total angular momentum states. Non-relativistic baryon states such as the proton and neutron have total angular momentum $j=\frac{1}{2}$ and magnetic quantum number, $m$. To explain our approach to incorporating spin, consider the appropriately symmetrized non-relativistic proton flavor-spin $j=\frac{1}{2}$, $m=\frac{1}{2}$ wave function:
\begin{eqnarray}
|P,+\rangle\equiv |uud\rangle \left(2|++-\rangle-|+-+\rangle-|-++\rangle\right)/(3\sqrt{2})+ {\rm cyclic\, permutations}.\label{proton}
\end{eqnarray}
The TF quark model is incapable of reproducing this linear combination. Instead given this wave function the TF model simply considers the probabilities of certain configurations determined by projections. In the proton world, the possible spin up configurations are:
\begin{eqnarray*}
(u^{\uparrow}u^{\uparrow}d^{\downarrow}),\quad (u^{\uparrow}u^{\downarrow}d^{\uparrow}).  \label{pconfig1}
\end{eqnarray*}
The proton is then said to be in the TF configuration:
\begin{eqnarray*}
\frac{2}{3}(u^{\uparrow}u^{\uparrow}d^{\downarrow})+\frac{1}{3}(u^{\uparrow}u^{\downarrow}d^{\uparrow}).   \label{pconfig2}
\end{eqnarray*}
(We assign no meaning to the flavor or spin sequential ordering.) We call this procedure the \lq\lq TF projection". It is the configurations which have distinct masses in the TF quark model. Thus, the TF spin model deals with probabilities of certain projected configurations rather than spin amplitudes. We assume the mass and other properties of the physical state are the probability weighted average of the configurations. Using the flavor degeneracy factors introduced in the last section, these two configurations are classified:
\begin{eqnarray*}
& (u^{\uparrow}u^{\uparrow}d^{\downarrow}):& \, g_1=1, N_1=2;\, g_2=1, N_2=1.\\ 
& (u^{\uparrow}u^{\downarrow}d^{\uparrow}):& \, g_1=3, N_1=1; N_2=0.  \label{pconfig3}
\end{eqnarray*}
That is, the first configuration has two identical particles (the $u$ quarks) and a second generalized flavor ($d$), whereas the second configuration simply has three non-identical but mass-degenerate quarks, with no second set of particle labelings necessary; note how the different spin $u$ quarks are treated as different flavors. We will have more to say on the violation of rotational symmetry later in this section.

Having introduced spin classifications into the model, we need to introduce a spin-splitting term. The classical interaction of a magnetically charged particle in an external magnetic field is given by
\begin{equation}
(H_{m}^{class})_{ij}=-{\vec m}_i\cdot {\vec B}_j,
\end{equation}
where ${\vec m}_i$ is the magnetic moment of particle $i$ and ${\vec B}_j$ is the external magnetic field of particle $j$. A magnetic dipole field is given by
\begin{equation}
{\vec B}^{d}=\frac{3\hat{r}(\hat{r}\cdot {\vec m})-{\vec m}}{r^3} +\frac{8\pi}{3}{\vec m}\,\delta({\vec r}),
\end{equation}
where $\delta({\vec r})$ is a Dirac delta function. In our case, the spherical symmetry of all integrations require the interaction to be of the form,
\begin{equation}
(H_{m}^{class})_{ij}=-\frac{8\pi}{3}{\vec m}_i\cdot {\vec m}_j\,\delta({\vec r}).
\end{equation}
These considerations lead us to postulate the form of the color magnetic interaction in our model:
\begin{equation}
(H_{m})_{ij}=-\frac{8\pi}{3}\gamma_i\gamma_j(S_z^{\cal I})_i(S_z^{\cal J})_j({\vec q}_i\cdot{\vec q}_j)\, \delta({\vec r}),\label{spininteraction}
\end{equation}
where the superscripts ${\cal I}$ and ${\cal F}$ are flavor labels and $i, j$ are particle number labels which will take on values from 1 to $N_{\cal I}$, $N_{\cal J}$. We define
\begin{equation}
\gamma_{i}\equiv \frac{\rm g}{2m_ic},
\end{equation}
where $\rm g$ is the color gyromagnetic factor, which takes on a value of 1 classically and a value of 2 to lowest order in QED perturbation theory for the electron and muon.

The problem of recovering a large enough spin-splitting in light mesons and baryons in non-relativistic models and even lattice QCD for large quark masses is well known. In this work, we will consider the g-factor to be an adjustable parameter due to relativistic and higher-order strong interaction effects. Other non-relativistic treatments keep the tree-level g-factor but instead introduce an adjustable wave function overlap factor\cite{dR} or use an extended interaction potential\cite{barnes}. Ultimately, the problem is only solved in a fully relativistic context at physical quark masses in lattice QCD\cite{lattice}. Here we hope to only reasonably model such effects. However, we expect spin effects to become smaller for increasing baryon number $A$, as will be seen in Section \ref{SSthree}.

Note that the specialization to spins along the z-axis in Eq.(\ref{spininteraction}) is appropriate for our individual spin basis. For two quarks in a colorless system,
\begin{equation}
\vec{q}_i\cdot\vec{q}_j=\left\{ \begin{array}{l} \frac{4}{3}g^2,  \,\,{\rm same\,\,color} \\
-\frac{2}{3}g^2,  \,\,{\rm different\,\,color} \end{array}\right.\label{color}
\end{equation}
(Note the notational distinction between \lq\lq$\rm g$" the color gyromagnetic factor and \lq\lq$g$" the color coupling constant and that $\alpha_s = g^2/(\hbar c)$.) The treatment of potential energy terms is explained in Ref.\cite{wilcox}. We sum over same flavor and different flavor contributions, using a single particle normalization for the particle densities ($\hat{n}^{\cal I} (r)= 3 n^{\cal I}(r)/N_{\cal I}$). We then average over the color interactions in Eq.(\ref{spininteraction}) using (\ref{color}) and the color probabilities, $P_{ij}$. The result for the magnetic spin interaction energy, $E_m$, may be written:
\begin{eqnarray}
E_m=\frac{8\pi}{3}\frac{9\times\frac{4}{3}g^2}{(3A-1)}\left[   \left(\frac{\hbar}{2}\right)^2 \sum_{\cal I}  \gamma_{\cal I}^2 \left( \frac{N_{\cal I}-1}{2N_{\cal I}}  \right)  \int d^3r\, \left(n^{\cal I}(r)\right)^2 \right. \nonumber \\
 +    \left. \sum_{{\cal I}\langle{\cal J}}   \gamma_{\cal I}  \gamma_{\cal J}  \sum_{i=1}^{N_{\cal I}}\sum_{j=1}^{N_{\cal J}}   (S_z^{\cal I})_i(S_z^{\cal J})_j  \int d^3r\, \left(   \frac{n^{\cal I}(r)n^{\cal J}(r)}{N_{\cal I}N_{\cal J}}  \right)  \right]. \label{spinenergy1}
\end{eqnarray}
Switching to the TF wave function, $f^{\cal I}(r)$ defined in Eq.(\ref{FinN2}), and using the $x$ variable defined in Eq.\ref{peq2} now gives
\begin{eqnarray}
E_m= \frac{16\hbar c}{3\pi^2 (3A-1)}\frac{(\frac{4}{3}\alpha_s)^4}{a}\left(\frac{\rm g}{2}\right)^2\left[   \sum_{\cal I}\alpha_{\cal I}^2\frac{N_{\cal I}-1}{2N_{\cal I}}  \int dx\, \frac{(f^{\cal I}(x))^3}{x}   \right. \nonumber\\
+ \left.  \sum_{{\cal I}\langle{\cal J}}   \alpha_{\cal I}  \alpha_{\cal J}  \sum_{i=1}^{N_{\cal I}}\sum_{j=1}^{N_{\cal J}} (\hat{S}_z^{\cal I})_i(\hat{S}_z^{\cal J})_j   \int dx \frac{1}{x} \frac{(f^{\cal I}(x))^{3/2}}{N_{\cal I}} \frac{(f^{\cal J}(x))^{3/2}}{N_{\cal J}} . \right]\label{spinenergy2}
\end{eqnarray}
where the unit normalized normalized spins have $(\hat{S}_z^{\cal J})_j = 1$ for spin up and $(\hat{S}_z^{\cal J})_j = -1$ for spin down along z.

As pointed out above, the introduction of spin labeling as an extended flavor attribute has the obvious shortcoming of loss of rotational symmetry. One way of stating the issue is that there is a conflict between the configurations used to determine flavor content, which use an individual spin basis with good values of $J^{\alpha}_z$, where $\alpha$ labels the particles, and the rotationally invariant total spin states which have good total angular momentum quantum numbers. This problem will not affect multi-quark total spin $0$ states, which are rotationally invariant, or spin $\frac{1}{2}$ states, whose $m= \pm \frac{1}{2}$ states have TF projections with the same mass. For example, the proton $m= \frac{1}{2}$ and $m= -\frac{1}{2}$ states project to the configurations $\frac{2}{3}(u^{\uparrow}u^{\uparrow}d^{\downarrow})+\frac{1}{3}(u^{\uparrow}u^{\downarrow}d^{\uparrow})$ and $\frac{2}{3}(u^{\downarrow}u^{\downarrow}d^{\uparrow})+\frac{1}{3}(u^{\uparrow}u^{\downarrow}d^{\downarrow})$, respectively, which have the same mass. However, the spin $\frac{3}{2}$ $m=\frac{1}{2}$ state,
\begin{eqnarray}
\frac{1}{\sqrt{3}}(|-++\rangle+|+-+\rangle+|++-\rangle), \nonumber
\end{eqnarray}
produces a different mass  than the $m=\frac{3}{2}$ state when projected into the individual spin basis states. The $m=\frac{1}{2}$ state corresponds to a $g_1=1, N_1=2; g_2=1, N_2=1$ configuration in our model, whereas the $m=\frac{3}{2}$ state corresponds to $g=1, N=3$. These have different masses.

\begin{table}
\caption{The Clebsh-Gordan coefficients $\langle j1;m,0|j1;j',m\rangle$}

\begin{center} 
\begin{tabular}{|c|c|c|c|c|c|c|}  \hline\hline
                   & $j=0$  &$j=1$     & $j=1$   & $j=2$ & $j=2$ & $j=3$\\
                   & $j'=0$ & $j'=0$    & $j'=2$   & $j'=1$ & $j'=3$ & $j'=2$\\ \hline
\vspace{-.3cm} &    	        & 	          &                 &              &              & \\
$m=3$ &  0    	        & 0  	          & 0                  &0                &0                &0\\ 
\vspace{-.3cm} &    	        & 	          &                 &              &              & \\ \hline
\vspace{-.3cm} &    	        & 	          &                 &              &              & \\
$m=2$ &  0     	        & 0                  & 0                  &0                &$\sqrt{\frac{5}{15}}$   &$-\sqrt{\frac{5}{21}}$  \\ 
\vspace{-.3cm} &    	        & 	          &                 &              &              & \\ \hline
\vspace{-.3cm} &    	        & 	          &                 &              &              & \\
$m=1$ &  0  	        & 0     	          &$\sqrt{\frac{3}{6}}$       &$-\sqrt{\frac{3}{10}}$  &$\sqrt{\frac{8}{15}}$   &$-\sqrt{\frac{8}{21}}$\\ 
\vspace{-.3cm} &    	        & 	          &                 &              &              & \\ \hline
\vspace{-.3cm} &    	        & 	          &                 &              &              & \\
$m=0$ & 1       &$-\sqrt{\frac{1}{3}}$               &$\sqrt{\frac{4}{6}}$            &$-\sqrt{\frac{4}{10}}$        &$\sqrt{\frac{9}{15}}$    &$-\sqrt{\frac{9}{21}}$      \\ 
\vspace{-.3cm} &    	        & 	          &                 &              &              & \\ \hline
\vspace{-.3cm} &    	        & 	          &                 &              &              & \\
$m=-1$ &  0  	        & 0     	          &$\sqrt{\frac{3}{6}}$       &$-\sqrt{\frac{3}{10}}$  &$\sqrt{\frac{8}{15}}$   &$-\sqrt{\frac{8}{21}}$\\ 
\vspace{-.3cm} &    	        & 	          &                 &              &              & \\ \hline
\vspace{-.3cm} &    	        & 	          &                 &              &              & \\
$m=-2$ &  0     	        & 0                  & 0                  &0                &$\sqrt{\frac{5}{15}}$   &$-\sqrt{\frac{5}{21}}$  \\ 
\vspace{-.3cm} &    	        & 	          &                 &              &              & \\ \hline
\vspace{-.3cm} &    	        & 	          &                 &              &              & \\
$m=-3$ &  0    	        & 0  	          & 0                  &0                &0                &0\\ 
 &    	        & 	          &                 &              &              & \\\hline
\hline

\end{tabular}
\end{center}
\label{CGtable}
\end{table}

We believe the best projection is always performed in a maximum $m$ states, $|j,m=j\rangle$. The reason has to do with what we will call the \lq\lq maximum compatibility" of the total spin and product spin basis states. As an example, let us consider the spin $|\frac{3}{2},\frac{3}{2}\rangle$ 3 quark states. Here there is only one way to build the state, namely $|\frac{1}{2}\rangle_1|\frac{1}{2}\rangle_2|\frac{1}{2}\rangle_3$ in the product spin basis. This state is also an eigenstate of total $\vec{J}^{\,2}$ and $J_z$. Thus we have ($\alpha=\{1,2,3\}$ is the particle label)
\begin{eqnarray}
 [\vec{J}^{\,2},J^{\alpha}_z]   |\frac{3}{2},\frac{3}{2}\rangle = 0,
\end{eqnarray}
for the commutator of $\vec{J}^{\,2}$ and the $J^{\alpha}_z$, expressing that the total and product quantum numbers are compatible. On the other hand the expectation value of the commutator of $\vec{J}^{\,2}$ with the individual $J^{\alpha}_z$ does not vanish, as evidenced by the nonzero matrix element:
\begin{eqnarray}
\langle \frac{1}{2},\frac{1}{2}|   [\vec{J}^{\,2},J^{\alpha}_z]   |\frac{3}{2},\frac{1}{2}\rangle = -3\hbar^2 \langle\frac{1}{2},\frac{1}{2}|J^{\alpha}_z|\frac{3}{2},\frac{1}{2}\rangle \\ \nonumber
= -3\hbar^2 \langle\frac{3}{2},1;\frac{1}{2},0|\frac{3}{2},1;\frac{1}{2},\frac{1}{2}\rangle\langle\alpha,\frac{1}{2}||\vec{J}||\alpha,\frac{3}{2}\rangle .
\end{eqnarray}
The first multiplicative factor on the right is the Clebsch-Gordan (\lq\lq CG") coefficient, 
\begin{equation}
\langle\frac{3}{2},1;\frac{1}{2},0|\frac{3}{2},1;\frac{1}{2},\frac{1}{2}\rangle=-\sqrt{\frac{1}{3}}, \nonumber
\end{equation}
\noindent and $\langle\alpha,\frac{1}{2}||\vec{J}||\alpha,\frac{3}{2}\rangle$ is the reduced matrix element. Clearly, the $j=3/2,m=3/2$ state is to be preferred over the $j=3/2,m=1/2$ state in the projection process. As a more general example, consider the $j=0,1,2,3$ states produced by coupling 6 quarks in zero angular momentum states. The $j=0$ states will have a rotationally invariant projection, whereas the $j=3,m=3$ state will be maximally compatible with the states with good individual $J^{\alpha}_z$ ($\alpha=\{1,2,3,4,5,6\}$). The other maximal $m$ states, $j=2,m=2$ and $j=1,m=1$ will also be the most compatible states to use with $j=2$ and $j=1$, although now the commutators with higher $j$ states will no longer vanish. This is evidenced by the CG coefficients in Table \ref{CGtable}. Note the zeros at the outside edge for the $(j=3,j'=2)$, $(j=2,j'=1)$ and $(j=1,j'=0)$ cases, expressing the fact that there is no lower $j'$ state to couple to for the highest $j,m$ value. In addition, the magnitude of the entries in each column are always smallest at the larger $|m|$ values, and largest for $m=0$ states. Again, we conclude the maximal $m$ states are to be preferred in the projection process.

\section{Ground State Baryon Fits}\label{Sfive}

In order for the model we are presenting to be predictive, we need to fix the phenomenological parameters. These 5 parameters are: \\
\begin{align*}
&B{\rm :\, \lq\lq Bag"\, constant}\\
&m_1{\rm :\, light\, quark\, mass}\\
&m_s{\rm : \,strange\, quark\, mass}\\
&\alpha_s{\rm :\, strong\, coupling\, constant}\\
&{\rm g: \,color\, gyromagnetic\, factor}\\
\end{align*}
First, we must understand that the model itself is not well designed to fit low quark number states, just as the atomic version would be poorly constructed to fit the low atomic binding energy states. In fact, applying this model to a \lq\lq gas" of three quarks would seem to be impossible. However, just like the atomic model, an unreasonable aspect of the mathematics of the TF model is that such a fit seems to be entirely reasonable. An aspect that helps is the fact that in the hadronic case there are both the octet and decuplet ground states energies available which may be used to help make the fit more robust; that is, there are more states available than parameters which need to be fit. 

There are eight types of base configurations and thirteen associated TF wave functions that are necessary to fit all the octet and decuplet states. Table \ref{picturetable} lists the wave functions present in the ground state hadrons as well as the particles which partly or wholly share that wave function. They are given in the spin \lq\lq up" state. In five cases there are two TF wave functions for each base configuration. The double subscript on these functions designates first the base configuration, and second the quark-type. The base configuration label runs from 1 through 8. The quark-type label takes on three possible values: light (\lq\lq $l$"), double light (\lq\lq $ll$") or strange (\lq\lq $s$"). For example, $f_{2,ll}$ is the configuration type in the second row of the Table associated with the double \lq\lq up" light sector, either $u^{\uparrow}u^{\uparrow}$ or $d^{\uparrow}d^{\uparrow}$. 

\begin{table}
\caption{TF configuration wave functions.}

\begin{center} 
\begin{tabular}{|c|c|c|c|c|c|c|c|c|}  \hline\hline
Base configuration (spin \lq\lq up")&Particle(s) &Wave function  &  $g_{1}$     & $N_{1}$   &$g_{2}$   &$N_{2}$      &$g_{s}$   &$N_{s}$\\ \hline
\vspace{-.3cm} &  &      &    &   &      &   &  &\\
$u^{\uparrow}u^{\downarrow}d^{\uparrow},d^{\uparrow}d^{\downarrow}u^{\uparrow}$ &$P,N$ &$f_{1}$  &3    &1   &--   &--   &-- &-- \\ 
\vspace{-.3cm} &  &      &    &   &      &   & & \\ \hline
\vspace{-.3cm} &  &      &    &   &      &   & & \\
$u^{\uparrow}u^{\uparrow}d^{\downarrow},u^{\uparrow}u^{\uparrow}d^{\uparrow},d^{\uparrow}d^{\uparrow}u^{\downarrow},d^{\uparrow}d^{\uparrow}u^{\uparrow}$ &$P,N,\Delta^{+}$ &$f_{2,ll},f_{2,l}$ &1 &2 &1 &$1^{+}$ &-- &-- \\ 
\vspace{-.3cm} &  &      &    &   &      &   & & \\ \hline
\vspace{-.3cm} &  &      &    &   &      &   & & \\
$u^{\uparrow}u^{\uparrow}u^{\uparrow},d^{\uparrow}d^{\uparrow}d^{\uparrow}$ &$\Delta^{++}$ &$f_{3}$ &1 &3 &-- &-- &-- &-- \\ 
\vspace{-.3cm} &  &      &    &   &      &   & & \\ \hline
\vspace{-.3cm} &  &      &    &   &      &   & & \\
$u^{\uparrow}u^{\downarrow}s^{\uparrow},d^{\uparrow}d^{\downarrow}s^{\uparrow}$ &  & & & & & & & \\
$u^{\uparrow}d^{\uparrow}s^{\downarrow},u^{\uparrow}d^{\downarrow}s^{\uparrow}$ &$\Sigma^{+},\Sigma^{0},\Sigma^{*0},\Lambda$ &$f_{4,l},f_{4,s}$ &2 &1 &-- &-- &$1^{+}$ &1 \\
$u^{\downarrow}d^{\uparrow}s^{\uparrow},u^{\uparrow}d^{\uparrow}s^{\uparrow}$ &  & & & & & & & \\ 
\vspace{-.3cm} &  &      &    &   &      &   & & \\ \hline
\vspace{-.3cm} &  &      &    &   &      &   & & \\
$u^{\uparrow}u^{\uparrow}s^{\downarrow},u^{\uparrow}u^{\uparrow}s^{\uparrow},d^{\uparrow}d^{\uparrow}s^{\downarrow},d^{\uparrow}d^{\uparrow}s^{\uparrow}$ &$\Sigma^{+},\Sigma^{*+}$ &$f_{5,l},f_{5,s}$ &1 &2 &-- &-- &1 &$1^{+}$ \\ 
\vspace{-.3cm} &  &      &    &   &      &   & & \\ \hline
\vspace{-.3cm} &  &      &    &   &      &   & & \\
$s^{\uparrow}s^{\downarrow}d^{\uparrow},s^{\uparrow}s^{\downarrow}d^{\uparrow}$ &$\Xi^{0}$ &$f_{6,l},f_{6,s}$ &$1^{+}$ &1 &-- &-- &2 &1 \\ 
\vspace{-.3cm} &  &      &    &   &      &   & & \\ \hline
\vspace{-.3cm} &  &      &    &   &      &   & & \\
$s^{\uparrow}s^{\uparrow}u^{\downarrow},s^{\uparrow}s^{\uparrow}u^{\uparrow},s^{\uparrow}s^{\uparrow}d^{\downarrow},s^{\uparrow}s^{\uparrow}d^{\uparrow}$ &$\Xi^{0},\Xi^{*0}$ &$f_{7,l},f_{7,s}$ &1 &$1^{+}$ &-- &-- &1 &2  \\ 
\vspace{-.3cm} &  &      &    &   &      &   & & \\ \hline
\vspace{-.3cm} &  &      &    &   &      &   & & \\
$s^{\uparrow}s^{\uparrow}s^{\uparrow}$ &$\Omega^{-}$ &$f_{8}$  &-- &-- &-- &-- &1 &3 \\ 
 &  &      &    &   &      &   & & \\ \hline \hline
\end{tabular}
\end{center}
\label{picturetable}
\end{table}

For the five configurations listed in Table \ref{picturetable} other than the three with either $g_1=3$ ($P,N$), $N_1=3$ ($\Delta^{++}$) or $N_s=3$ ($\Omega^-$), the consistency condition, Eq.(\ref{consistent}), does not formally have a solution when all the $g$ or $N$ values are integer. For each of the five configurations this occurs when the ($N_1, N_2$), ($N_1, N_s$) or ($g_1, g_s$) sector takes on a value of either (2,1) or (1,2). These cases are indicated in the Table with the designation $1^+$ to indicate that the solution to (\ref{consistent}) is defined by an infinitesimal approach to that parameter from above. For example, in the case of the $f_{2,ll},f_{2,l}$ wave functions, which contribute to the energies of the $P$, $N$ and $\Delta^+$ states, we have the robust numerical condition
\begin{equation}
f_{2,ll}(x)=1.23162 f_{2,l}(x),\label{f1,f2}
\end{equation}
from the consistency condition, Eq.(\ref{consistent}), for the relationship between $f^1(x)$ ($f_{2,ll}(x)$) and $f^2(x)$ ($f_{2,l}(x)$) for $0< x< x_2$ in the limit $N_2\longrightarrow 1^+$. We use this limit to {\it define} the value of the model for  $N_2=1$ exactly. (Note that $f_{2l}(x)=0$ for $x_2< x< x_1$ since it is the TF wave function with the smaller radius.)

Table \ref{picturetable2} gives both the base particle configuration, to be used in conjunction with Table \ref{picturetable}, as well as the actual formulas for the spin splittings from Eq.(\ref{spinenergy2}) above. Besides the proton and neutron, note that a number of other particles share the same base particle energies before magnetic spin splitting terms are added because we assume the $u$ and $d$ quark masses are degenerate. These degeneracies are: ($\Sigma^{+}, \Sigma^-$), ($\Xi^{0}, \Xi^-$), ($\Delta^{++}, \Delta^-$), ($\Delta^{+}, \Delta^0$), ($\Sigma^{*+}, \Sigma^{*-}$), ($\Xi^{*0}, \Xi^{*-}$), and the ($\Lambda$, $\Sigma^0$, $\Sigma^{*0}$) particles. We list only the $P$, $\Sigma^{+}$, $\Xi^0$, $\Delta^{++}$, $\Delta^{+}$, $\Sigma^{*+}$, and $\Xi^{*0}$ particles in Table \ref{picturetable2} because of these degeneracies. The $\Lambda$, $\Sigma^0$ and $\Sigma^{*0}$ all get different masses after spin interaction and are listed separately.

\begin{table}
\caption{Base particle asignments and spin energy contributions of ground state baryons. The various $f$ functions used here are from Table \ref{picturetable}.}

\begin{center} 
\begin{tabular}{|c|c|}  \hline\hline
           &   \\
        
Base particle configuration & Magnetic Energy Splitting ($C \equiv \frac{16 \hbar c}{3 \pi^{2}(3A-1)}\frac{(\frac{4}{3}\alpha_{s})^{4}}{a}(\frac{g}{2})^{2}$) \\
 (spin  \lq\lq up", maximal $m$) &            \\ \hline
 \vspace{-.3cm} & \\
$P=\frac{1}{3}\left(u^{\uparrow}u^{\downarrow}d^{\uparrow}\right)+\frac{2}{3}\left(u^{\uparrow}u^{\uparrow}d^{\downarrow}\right)$ &$E^{P}_{m}=C\left\{\frac{1}{3}\left(-\int dx\frac{(f_{1})^{3}}{x}\right)+\frac{2}{3}\left(\frac{1}{4}\int dx\frac{(f_{2,ll})^{3}}{x}-\int dx\frac{(f_{2,ll}f_{2,l})^{3/2}}{x}\right)\right\}$ \\ 
  \vspace{-.3cm} & \\ \hline
   \vspace{-.3cm} & \\
$\Lambda=u^{\uparrow}d^{\downarrow}s^{\uparrow}$ &$E^{\Lambda}_{m}=C\left\{-\int dx \frac{(f_{4,l})^{3}}{x}\right\}$ \\
   \vspace{-.3cm}& \\ \hline
    \vspace{-.3cm} & \\
$\Sigma^{+}=\frac{1}{3}\left(u^{\uparrow}u^{\downarrow}s^{\uparrow}\right)+\frac{2}{3}\left(u^{\uparrow}u^{\uparrow}s^{\downarrow}\right)$ &$E^{\Sigma^{+}}_{m}=C\left\{\frac{1}{3}\left(-\int dx\frac{(f_{4,l})^{3}}{x}\right)+\frac{2}{3}\left(\frac{1}{4}\int dx\frac{(f_{5,l})^{3}}{x}-\alpha_{str} \int dx\frac{(f_{5,l}f_{5,s})^{3/2}}{x}\right)\right\}$ \\ 
   \vspace{-.3cm}& \\ \hline
     \vspace{-.3cm}& \\
$\Sigma^{0}=\frac{1}{3}\left(u^{\uparrow}d^{\downarrow}s^{\uparrow}\right)+\frac{2}{3}\left(u^{\uparrow}d^{\uparrow}s^{\downarrow}\right)$ &$E^{\Sigma^{0}}_{m}=C\left\{\frac{1}{3}\left(-\int dx\frac{(f_{4,l})^{3}}{x}\right)+\frac{2}{3}\left(\int dx\frac{(f_{4,l})^{3}}{x}-2\alpha_{str} \int dx\frac{(f_{4,l}f_{4,s})^{3/2}}{x}\right)\right\}$ \\ 
  \vspace{-.3cm} & \\ \hline
     \vspace{-.3cm}& \\
$\Xi^{0}=\frac{1}{3}\left(s^{\uparrow}s^{\downarrow}u^{\uparrow}\right)+\frac{2}{3}\left(s^{\uparrow}s^{\uparrow}u^{\downarrow}\right)$ &$E^{\Xi^{0}}_{m}=C\left\{\frac{1}{3}\left(-\alpha^{2}_{str}\int dx\frac{(f_{6,s})^{3}}{x}\right)+\frac{2}{3}\left(\frac{1}{4}\alpha^{2}_{str}\int dx\frac{(f_{7,s})^{3}}{x}-\alpha_{str} \int dx\frac{(f_{7,l}f_{7,s})^{3/2}}{x}\right)\right\}$ \\ 
   \vspace{-.3cm}& \\ \hline
    \vspace{-.3cm} & \\
$\Delta^{++}=u^{\uparrow}u^{\uparrow}u^{\uparrow}$ &$E^{\Delta^{++}}_{m}=C\left\{\frac{1}{3}\int dx \frac{(f_{3})^{3}}{x}\right\}$ \\ 
   \vspace{-.3cm}& \\ \hline
     \vspace{-.3cm}& \\
$\Delta^{+}=u^{\uparrow}u^{\uparrow}d^{\uparrow}$ &$E^{\Delta^{+}}_{m}=C\left\{\frac{1}{4}\left(\int dx\frac{(f_{2,ll})^{3}}{x}\right)+\left(\int dx\frac{(f_{2,ll}f_{2,l})^{3/2}}{x}\right)\right\}$ \\ 
   \vspace{-.3cm}& \\ \hline
    \vspace{-.3cm} & \\
$\Sigma^{*+}=u^{\uparrow}u^{\uparrow}s^{\uparrow}$ &$E^{\Sigma^{*+}}_{m}=C\left\{\frac{1}{4}\left(\int dx\frac{(f_{5,l})^{3}}{x}\right)+\alpha_{str}\left(\int dx\frac{(f_{5,l}f_{5,s})^{3/2}}{x}\right)\right\}$ \\ 
   \vspace{-.3cm}& \\ \hline
     \vspace{-.3cm}& \\
$\Sigma^{*0}=u^{\uparrow}d^{\uparrow}s^{\uparrow}$ &$E^{\Sigma^{*0}}_{m}=C\left\{\left(\int dx\frac{(f_{4,l})^{3}}{x}\right)+2\alpha_{str}\left(\int dx\frac{(_{4,l}f_{4,s})^{3/2}}{x}\right)\right\}$ \\ 
   \vspace{-.3cm}& \\ \hline
     \vspace{-.3cm}& \\
$\Xi^{*0}=u^{\uparrow}s^{\uparrow}s^{\uparrow}$ &$E^{\Xi^{*0}}_{m}=C\left\{\frac{1}{4}\alpha^{2}_{str}\left(\int dx\frac{(f_{7,s})^{3}}{x}\right)+\alpha_{str}\left(\int dx\frac{(f_{7,l}f_{7,s})^{3/2}}{x}\right)\right\}$ \\ 
  \vspace{-.3cm} & \\ \hline
     \vspace{-.3cm}& \\
$\Omega^{-}=s^{\uparrow}s^{\uparrow}s^{\uparrow}$ &$E^{\Omega^{-}}_{m}=C\left\{\frac{1}{3}\int dx \frac{(f_{8})^{3}}{x}\right\}$ \\ 
& \\ \hline\hline
\end{tabular}
\end{center}
\label{picturetable2}
\end{table}

There is a subtlety concerning the $\Xi^{0,-}$ and $\Xi^{*0,-}$ particles which prevents us from including them in our numerical results. All these particles contain two strange quarks and a light quark. In our numerical simulations we find that the two strange quark wave functions, $f_{6,s}$ or $f_{7,s}$, have a smaller radius than the single light quark wave functions, $f_{6,l}$ or $f_{7,l}$, which in Table \ref{picturetable} are characterized as either $g_1=1^+, N_1=1$ or $g_1=1, N_1=1^+$. This implies that the right hand side of Eq.(\ref{one}) is zero for the region $x_2< x< x_1$, where $f^2(x)=0$, and that the light quark is \lq\lq free" with no restoring force. The equations can still be formally solved and the normalization condition for $f^1$ in Eq.(\ref{five}) can still be fulfilled, but we reject this solution as unrealistic. Thus, although the $\Xi^{0,-}$ and $\Xi^{*0,-}$ wave functions are formally calculable, we don't believe the model can give correct wave functions in this case, and they are omitted from our numerical results. This is a case of a gas of a single particle not being properly calculable.

A computer program in {\it Mathematica} has been developed and the parameters which fit the low energy baryon spectrum have been determined by explicit numerical energy minimization. The first step in this process is the numerical solution of the $f^2(x)$ (inner) function differential equation, Eqs.(\ref{18}) or (\ref{19}), beginning with a guess for the initial slope, followed by the reconstruction/solution of the $f^1(x)$ (outer) function for a given set of external parameters and normalization conditions. Once this is done for all particle states, a chi-square minimization is carried out among the hadron masses. The values assumed for the experimental masses in Table \ref{table3} are rounded to the nearest $MeV$. In addition in Table \ref{table3}, the $\Sigma^+$ row actually lists the average experimental mass of the $\Sigma^+$ and $\Sigma^-$, the $\Sigma^{*+}$ row actually lists the average experimental mass of the $\Sigma^{*+}$ and $\Sigma^{*-}$, and a nominal value of 1232 $MeV/c^2$ is used for all the $\Delta$ particles. Using a grid search algorithm we have found a best fit with the values: $B^{1/4}=84.4$ $MeV$, $m_1=290$ $MeV/c^2$, $m_s= 507$ $MeV/c^2$ (corresponding to $\alpha_{str}=0.572$ as the mass ratio in Eq.(\ref{alpha})), $\alpha_s=0.430$, $\rm g=7.09$. Here we are are fitting nine data points with five parameters. The spin splitting terms are treated as perturbations and are not included in the numerical energy minimizations to simplify the calculations. In addition, the color gyromagnetic value, g, was determined as a separate chi-squared minimization for each set of $B, m_1, m_s, \alpha_s$ values. These values are phenomenologically appropriate and reasonable. The bag constant is lower than in the original application\cite{bag}, which means the non-relativistic fit is actually more self-consistent. The strong coupling constant is consistent with a bag model designed for heavy-light systems where the center of mass motion less of an issue\cite{wilcox2}. The color gyromagnetic factor, $\rm g$, is large compared to the electrodynamic case for electrons ($\sim 2$), but not unreasonably so.

\begin{table}
\caption{The full set of calculable octet and decuplet particle energies in the TF quark model. Particles which have degenerate mass in the $m_u=m_d$ limit are not listed.}

\begin{center} 
\begin{tabular}{|c|c|c|c|c|c|c|c|c|}  \hline\hline
Particle            &Bag Radius             &Rest Mass              & T            &U                      &Spin                     &Volume                 &Total                   &Exp  \\ 
            & $(fm)$          &($MeV/c^2$)    & ($MeV/c^2$)     &($MeV/c^2$)      &($MeV/c^2$)         & ($MeV/c^2$)       &($MeV/c^2$)          &($MeV/c^2$)\\ \hline
\vspace{-.3cm}&          &        &      &         &           &       &           & \\
$P$   &1.48 (1.37) &870 &219.6 &-145.3 &-70.9 &83.7 &957.1 &939 \\ 
\vspace{-.3cm}&          &        &      &         &           &       &           & \\ \hline
\vspace{-.3cm}&          &        &      &         &           &       &           & \\
$\Delta^{+}$ &1.48 &870 &237.6 &-140.3 &151.4 &89.6 &1208 &1232 \\ 
\vspace{-.3cm}&          &        &      &         &           &       &           & \\ \hline
\vspace{-.3cm}&          &        &      &         &           &       &           & \\
$\Delta^{++}$ &1.68 &870 &250.6 &-124.0 &104.5 &132.2 &1233 &1232 \\ 
\vspace{-.3cm}&          &        &      &         &           &       &           & \\ \hline
\vspace{-.3cm}&          &        &      &         &           &       &           & \\
$\Lambda$ &1.29 &1087 &200.0 &-151.9 &-81.0 &59.5 &1114 &1116 \\ 
\vspace{-.3cm}&          &        &      &         &           &       &           & \\ \hline
\vspace{-.3cm}&          &        &      &         &           &       &           & \\
$\Sigma^{+}$ &1.40 (1.29) &1087 &233.2 &-152.4 &-41.3 &70.9 &1197 &1193 \\ 
\vspace{-.3cm}&          &        &      &         &           &       &           & \\ \hline
\vspace{-.3cm}&          &        &      &         &           &       &           & \\
$\Sigma^{0}$ &1.29 &1087 &200.0 &-151.9 &-43.7 &59.5 &1151 &1193 \\
\vspace{-.3cm}&          &        &      &         &           &       &           & \\ \hline
\vspace{-.3cm}&          &        &      &         &           &       &           & \\
$\Sigma^{*+}$ &1.40 &1087 &249.8 &-152.7 &144.3 &76.6 &1405 &1385 \\ 
\vspace{-.3cm}&          &        &      &         &           &       &           & \\ \hline
\vspace{-.3cm}&          &        &      &         &           &       &           & \\
$\Sigma^{*0}$ &1.29 &1087 &200.0 &-151.9 &187.1 &59.5 &1382 &1384 \\
\vspace{-.3cm}&          &        &      &         &           &       &           & \\ \hline
\vspace{-.3cm}&          &        &      &         &           &       &           & \\
$\Omega^{-}$ &1.45 &1521 &196.0 &-146.5 &31.3 &83.9 &1686 &1672 \\  
&          &        &      &         &           &       &           & \\ \hline\hline
\end{tabular}
\end{center}
\label{table3}
\end{table}

\begin{figure}
\begin{center}
\leavevmode
\includegraphics*[trim=00 000 0 0, clip,scale=1.0]{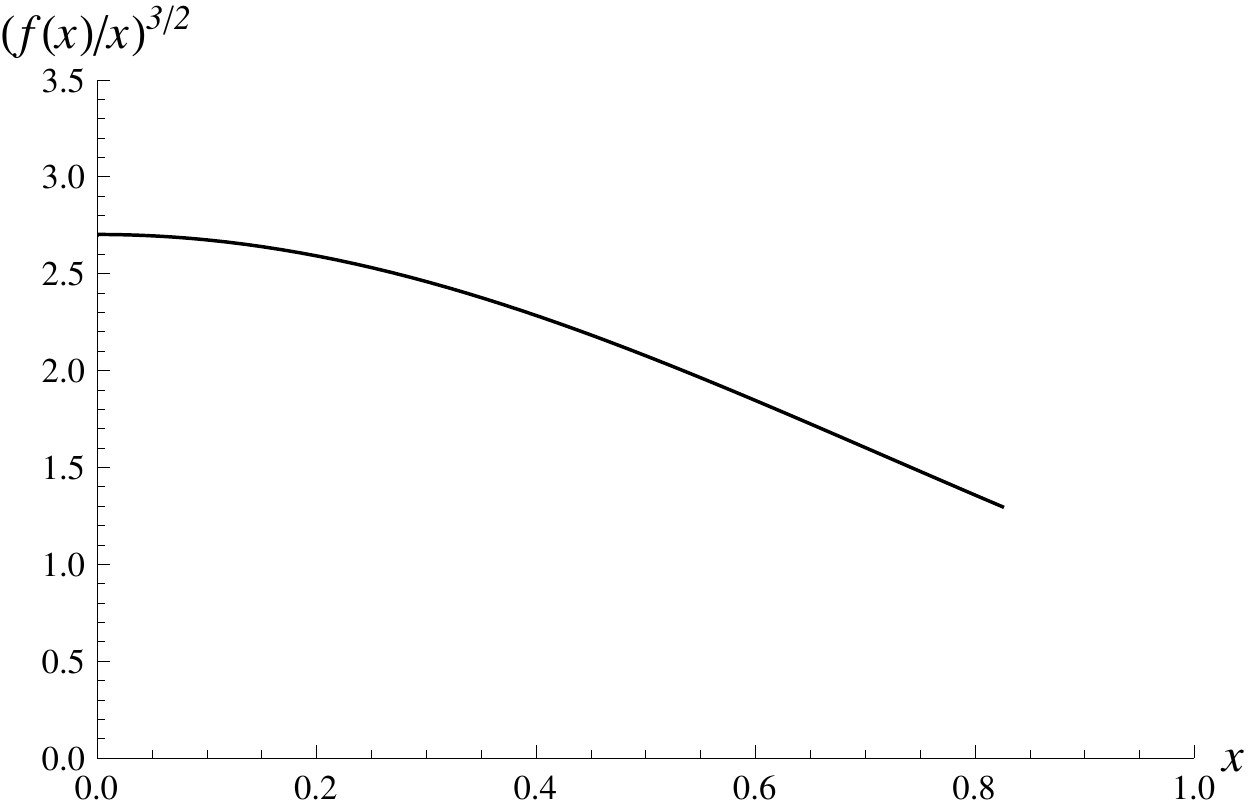}
\caption{The density profile of the $g=3, N=1$ proton or neutron TF wave function in terms of the dimensionless variable $x$.}
\label{fig1}
\vspace{2cm}
\includegraphics*[trim=00 000 0 0, clip,scale=1.0]{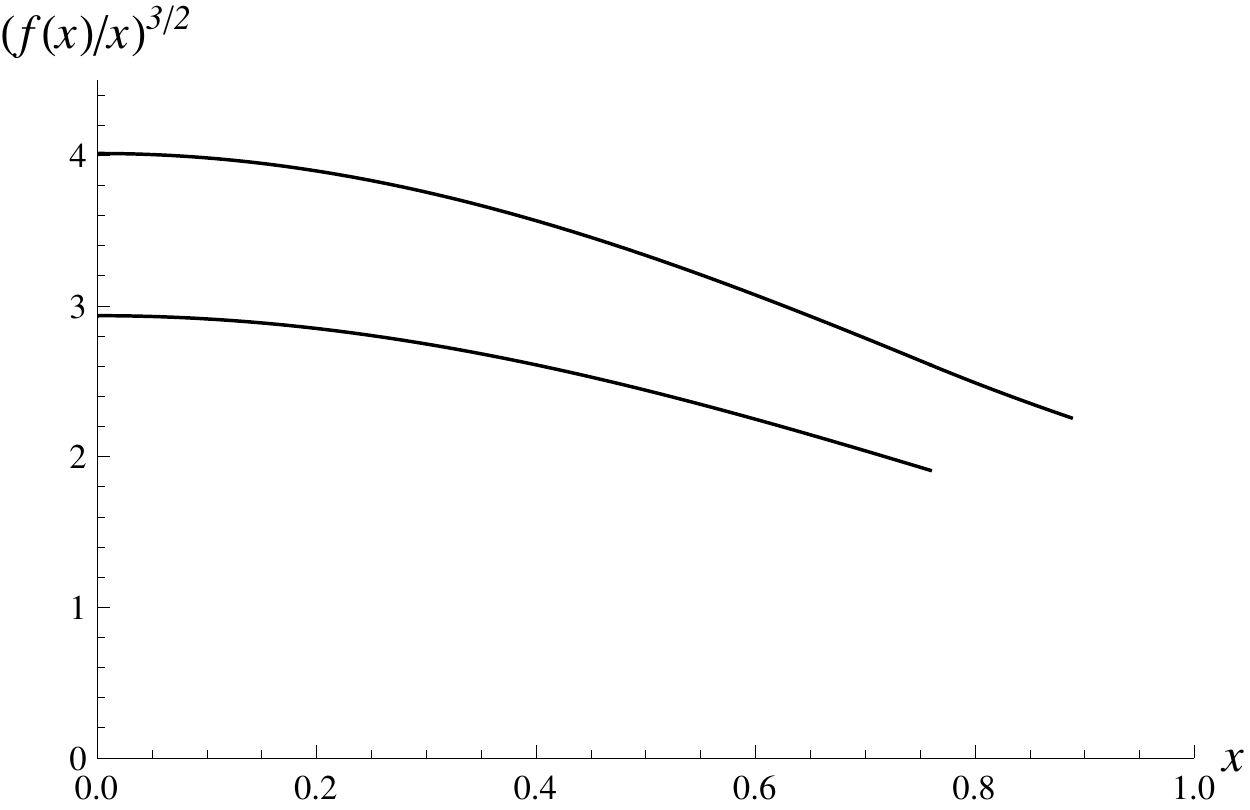}
\caption{The density profile of the $N_1=2,g_1=1;N_2=1,g_2=1 $ part of the proton or neutron TF wave function in terms of the dimensionless variable $x$. The top $f$ function represents $f_{2,ll}$ and the bottom represents $f_{2,l}$ from Table \ref{picturetable}. These wave functions are also relevant to the $\Delta^{+,0}$ particles.}
\label{fig2}
\end{center}
\end{figure}

Table \ref{table3} gives a full accounting of the various particle energies in our model. Notice that the final particle energies involved are consistent with the non-relativistic assumption partly because there are large cancellations between the kinetic and potential energies. One can see that the model overestimates the strong isospin breaking effects from the particle wave functions for the differently charged $\Delta$, $\Sigma$ and $\Sigma^*$ particles.  We note that the $\Delta^{++,-}$ and $\Delta^{+,0}$ particles actually originate in different base configurations; similarly for the $\Sigma^{+,-}$ and $\Sigma^{0}$ as well as the $\Sigma^{*+,-}$ and $\Sigma^{*0}$. There should actually be such isospin breaking effects in nature from the exclusion principle, but clearly they are not as large as seen here. As far as the authors know, lattice QCD data has not been examined for such effects. Although the fit is not as good as for a typical non-relativistic quark model, we would not expect it to be. As explained above, the known ground state baryons badly violate the many-quark semi-classical assumption. Nevertheless, we are encouraged by the overall reasonableness of the fit.

Some states will be well represented while others will not. It is especially encouraging that the masses of states such as the $\Lambda$, $\Omega$ and $\Delta^{++}$ seem to be accurate. These states are \lq\lq purest" in terms of their lack of mixings. The $\Lambda$ is the only baryon built out of a partially anti-symmetric flavor-spin combination, and the $\Omega$ and $\Delta^{++}$ have especially simple unmixed flavor-spin wave functions. The $\Lambda$ is especially important because of the possibility of strange matter formation, which will be studied in Sections \ref{SSone} and \ref{SSthree}.

We think it is informative to compare the ground-state baryon mass fit achieved here with two historically important hadronic models. The first column of Table \ref{table6} lists the total TF quark model masses again from Table \ref{table3}, the second column gives the original fit of the MIT bag model\cite{bag} and the third column gives the fit from the Isgur and Capstick relativized quark model\cite{Isgur}. (Note that the $\Xi$ and $\Xi^{*}$ TF masses have not been listed in Table \ref{table6} for the reason pointed out earlier in this section.) 
We evaluate the average absolute mass difference between the computed masses and the experimental masses as a rough measure of goodness of fit. The bag model average mass difference is 10.8 $MeV/c^2$ while the Isgur/Capstick model average mass difference is 13.2 $MeV/c^2$. The isospin splittings for the TF case makes a direct comparison impossible. However, if the isospin states are first averaged before the mass difference is calculated, we obtain 12.2 $MeV/c^2$ for the overall average. On the other hand, taking the average of the separate high and low mass differences gives 13.5 $MeV/c^2$. In this sense the goodness of fit of the TF model is approximately the same as the other two models.

\begin{table}
\caption{Comparison among different models. Various charge states are identified with superscripts for the present TF quark model. The experimental results represent averages over the charge states.}

\begin{center} 
\begin{tabular}{|c|c|c|c|c|}  \hline\hline
Particle            &TF quark model             &MIT Bag model\cite{bag}              & Relativized model\cite{Isgur}         &\quad Exp  \quad       \\ 
                        & ($MeV/c^2$)                 &($MeV/c^2$)                            & ($MeV/c^2$)                                       &\quad($MeV/c^2$)  \quad    \\ \hline
\vspace{-.3cm}&                          &                                      &                                                 &        \\
$P,N$                &957                     &938                               &960                                           & 939 \\ 
\vspace{-.3cm}&                           &                                     &                                                 &         \\ \hline
\vspace{-.3cm}&                           &                                     &                                                 &         \\
$\Delta$           &$1233^{++,-}, 1208^{+,0}$                   &1233                             &1230                                                 &1232          \\ 
\vspace{-.3cm}&                           &                                     &                                                 &          \\ \hline
\vspace{-.3cm}&                           &                                     &                                                 &         \\
$\Lambda$       & 1114                   &1105                            &1115                                          &1116           \\ 
\vspace{-.3cm}&                            &                                    &                                                  &         \\ \hline
\vspace{-.3cm}&                            &                                    &                                                  &         \\
$\Sigma$  &$1197^{+,-}, 1151^{0}$                    &1144                            &1190                                          &1193       \\ 
\vspace{-.3cm}&                            &                                    &                                                  &         \\ \hline
\vspace{-.3cm}&                            &                                    &                                                  &          \\

$\Xi           $  &       -             &    1289                         &     1305                                      &  1318      \\ 
\vspace{-.3cm}&                            &                                    &                                                  &         \\ \hline
\vspace{-.3cm}&                            &                                    &                                                  &          \\

$\Sigma^{*}$  &$1405^{+,-},  1382^{0} $                &1382                             &1370                                          & 1385         \\ 
\vspace{-.3cm}&                           &                                     &                                                  &           \\ \hline
\vspace{-.3cm}&                           &                                     &                                                  &           \\

$\Xi^{*}$  &   -                &    1529                         &     1505                                     & 1533       \\ 
\vspace{-.3cm}&                           &                                     &                                                  &           \\ \hline
\vspace{-.3cm}&                           &                                     &                                                  &           \\
$\Omega$ &1686               & 1672                            &1635                                          &1672            \\[1ex]     \hline
                                                 
\end{tabular}
\end{center}
\label{table6}
\end{table}

The radii listed in Table \ref{table3} are actually the bag radii of the wave functions. In two cases, the $P$ and $\Sigma^+$, two radii are given since there are two configurations involved; see Table \ref{picturetable2}. The larger radius in both cases is associated with the configuration with the larger particle number $N_1$, which is exactly as one would expect from the exclusion principle. As an example of the form of the TF wave functions, Figures \ref{fig1} and \ref{fig2} show the two parts of the proton wave function. Note that $x=1$ corresponds to a distance of 1.67\,$fm$ from Eq.(\ref{peq2}). Also notice an unusual aspect of our model due to its statistical nature: there is a clean separation of quark flavors phases for ($u,d$) quarks in the outer part of the $\Delta^{+,0}$ TF wave functions (see Fig.\ref{fig2}), as well as the (light, strange) sectors in the $\Sigma^{*+,-}$ particles. (This would be true as well for the $\Xi^{*0,-}$ particles had they been constructed.)

\begin{table}
\caption{Calculated baryonic electromagnetic squared charge radii (in $fm^2$) compared to various result and models.}

\begin{center} 
\begin{tabular}{|c|c|c|c|c|c|}  \hline\hline
{\bf Particle}            &{\bf TF model}     &   {\bf lattice}\cite{Leinweber}  &  {\bf HB$\chi$PT}\cite{ramsey}   &  {\bf Chiral}\cite{berger}    &  {\bf Expt.}\cite{PDG}   \\  \hline
$P$   &1.18 &  0.685$\pm$ 0.066   &   0.735   &    0.82   &  0.769$\pm$ 0.009\\ \hline
$N$   & -0.11    &  -0.158$\pm$ 0.033     &    -0.113   &   -0.13     &  -0.1161$\pm$ 0.0022\\  \hline
$\Delta^{++}$ &1.62 &   -  &   -   &  0.43      &  -\\ \hline
$\Delta^{+}$ &1.25 &   -  &   -   &    0.43    &  -\\ \hline
$\Delta^{0}$     &-0.17 & -   &    -    &   0.0    & - \\ \hline 
$\Delta^{-}$  &1.62   &   - &  -    &   0.43    &   -\\ \hline
$\Lambda$ &0.11  &  0.010$\pm0.009$  &   -0.284   &    0.03    &  -\\ \hline
$\Sigma^{+}$ &1.21  &  0.749$\pm$ 0.072  &  1.366    &   1.13     & - \\ \hline
$\Sigma^{-}$  &0.90 &  0.657$\pm$ 0.058 &     &   0.72     &  0.61$\pm$ 0.15 \\ \hline
$\Sigma^{0}$ &0.12 &  - &   -   &    0.20   &  - \\ \hline
$\Sigma^{*+}$ &1.27 &  - &  0.798    &   0.42     &  -\\ \hline
$\Sigma^{*-}$ &0.93 &   - &   -   &   0.37    &  - \\ \hline
$\Sigma^{*0}$ &0.11 &   - &   -   &  0.03      &  -\\ \hline
$\Omega^{-}$ &1.16  &  - &   -   &    0.29     & - \\ \hline
\hline
\end{tabular}
\end{center}
\label{table5}
\end{table}

To obtain radii comparable to experiment, we also calculate the electromagnetic squared charge radii from our charge distributions in Table \ref{table5}. We calculate the squared charge radius for a particle $P$ with total charge $Q$ from
\begin{equation}
\langle r^2\rangle_P = \frac{R^2}{Q} \sum_{B,q}  P_B Q_q \langle x^2\rangle_{B,q}.\label{rsquared}
\end{equation}
The sum is over base configurations ($B=1,\dots ,8$) and quark types ($q=l,ll,s$) in Table \ref{picturetable}. $R$ is given by Eq.(\ref{peq2}), $P_B$ are configuration  weightings read off from Table \ref{picturetable}, and $Q_q$ are the individual quark charges in units of the magnitude of the electron charge. When the total charge of the particle is zero, we must replace $Q$ with 1. We also have from Eq.(\ref{norm})
\begin{equation}
\langle x^2\rangle_{B,q} = \frac{3A}{N_{q}}\int_0^{x_{max}} dx\, x^{5/2} (f_{B,q}(x))^{3/2}.
\end{equation}
Then for the proton for example,
\begin{equation}
\langle r^2\rangle_P = R^2\left(\frac{1}{3}\langle x^2\rangle_1 +\frac{2}{3}\left(\frac{4}{3}\langle x^2\rangle_{2,ll}-\frac{1}{3}\langle x^2\rangle_{2,l}\right)\right),\label{r2proton}
\end{equation}
whereas for the neutron
\begin{equation}
\langle r^2\rangle_N = R^2\frac{2}{3}\left(-\frac{2}{3}\langle x^2\rangle_{2,ll}+\frac{2}{3}\langle x^2\rangle_{2,l}\right).\label{r2neutron}
\end{equation}
The $\langle x^2\rangle_1$ term corresponding to the $B=1$ configuration is absent here since the same function weights both the positive and negative charges.

The Table \ref{table5} radii are relatively large compared, for example, to measured charged particle electromagnetic radii. This is a result of the fit with a smaller value of $B^{1/4}$ than most standard bag model phenomenologies. For comparison, we have also listed in this Table recent results from a lattice calculation\cite{Leinweber}, heavy baryon chiral perturbation theory\cite{ramsey}, a chiral constituent quark model\cite{berger}, and the three known results from experiment\cite{PDG}. (Note that Ref.\cite{bag} gives the charge radii squared of the proton and neutron as 0.53\,$fm^2$ and 0, respectively. Ref.\cite{Isgur} did not calculate this quantity.) Our proton and $\Sigma^-$ squared charge radii are too large. However, the more extended $d$ quark TF wave function seen in Fig.\ref{fig2} results in a negative squared charge radius value, comparable to experiment.

We now turn to try applying the phenomenology developed here to quark configurations in a number of sectors.

\section{Three Case Studies}\label{Ssix}

In this section we will apply the fit found in the previous section to begin to explore some of the interesting phenomenology of high number multi-quark states. In particular, we will examine three possibilities: the  {\it H}-dibaryon, high multi-quark strange states, and nucleon-nucleon 6 quark resonances.

\subsection{{\it H}-Dibaryon Considerations}\label{SSone}
The {\it H}-dibaryon, which is an isospin $I=0$, total angular momentum $J=0$ $uuddss$ flavor state, is an interesting application of the model. Although there are many lattice results\cite{detmold,inoue,sakai,wetzorke,luo,inuoe2,beane,beane2,inuoe3,inuoe4}, there are systematic issues in the calculations due to lattice volume and quark mass effects. There are experimental results\cite{takahashi,yoon} which limit the bound state energy to shallow values if indeed it exists at all\cite{trattner}. The lattice results must be extrapolated in quark mass, and the results of this procedure are still rather uncertain\cite{detmold}. Our model is designed to evaluate states for further lattice study, so it is interesting to study this possible bound state with the new tool we now possess, even though we are doing this study \lq\lq post-lattice" at this point.

We may form a total $I=0$, $j=0$ combination from combining two particles in the ($\Lambda$$\Lambda$), ($\Sigma$$\Sigma$) or ($\Xi N$) systems\cite{meissner}. The individual particle isospins are $I_{\Lambda}=0$, $I_{\Sigma}=1$, $I_{N}=\frac{1}{2}$ and $I_{\Xi}=\frac{1}{2}$. One can look up the Clebsch-Gordon coefficients to couple these systems to a $|I_{tot}=0\rangle$ state. In quantum mechanics the states mix with one another, giving a coupled channel problem\cite{inoue}. In principle, one would proceed here exactly as one would for the quantum mechanical problem; assume a general linear combination and minimize with respect to the parameters to find the ground state. Such considerations are beyond the present application purview. However, given the quantum wave functions each of these combinations may be examined separately for the lowest mass.

The possible TF quark 4 quark $uuddss$ $j=0$ combinations are only three in number (we don't list some redundant configurations in the $m_u=m_d$ isospin limit):
\begin{eqnarray}
\lefteqn{\quad(u^{\uparrow}u^{\uparrow}d^{\downarrow}d^{\downarrow}s^{\uparrow}s^{\downarrow}): N_1=2, g_1=2,=2; N_s=1,g_s=2,} \nonumber \\
&&(u^{\uparrow}u^{\uparrow}d^{\uparrow}d^{\downarrow}s^{\uparrow}s^{\uparrow}): N_1=2, g_1=1; N_2=1, g_2=2;  N_s=2, g_s=1,\nonumber \\
&&(u^{\uparrow}u^{\downarrow}d^{\uparrow}d^{\downarrow}s^{\uparrow}s^{\downarrow}): N_1=1, g_1=4; N_s=1, g_s=2.\nonumber
\end{eqnarray}
Unfortunately, the $u, d$ asymmetric second combination, $(u^{\uparrow}u^{\uparrow}d^{\uparrow}d^{\downarrow}s^{\uparrow}s^{\uparrow})$, can not be formed here because it involves {\it three} separate and different quark wave functions. This limits us to investigation of the $\Lambda\Lambda$ state only, although this is also likely the lightest state from Pauli blocking considerations. We emphasize that the technical limitation here is simply due to the software developed so far, not in the model itself.

When fully reduced, the bare TF configuration for the $\Lambda\Lambda$ $I=0,j=0$ state is given by:
\begin{equation*}
\frac{1}{2}(u^{\uparrow}u^{\uparrow}d^{\downarrow}d^{\downarrow}s^{\uparrow}s^{\downarrow}) +\frac{1}{2}(u^{\uparrow}u^{\downarrow}d^{\uparrow}d^{\downarrow}s^{\uparrow}s^{\downarrow}).
\end{equation*}
The energy spin term is also easily calculable and is included in our results. Our result for the $\Lambda\Lambda$ state is 2228 MeV/c$^2$. The component energies and bag radius (not charge radius) are given in Table \ref{multiquark}. This is 86 MeV/c$^2$ more than the two $\Lambda$ threshold. Our present results indicate the {\it H}-dibaryon state is not bound in the TF quark model.

\subsection{High Multi-Quark Strange States}\label{SSthree}

We have studied the possibility of a bound {\it H}-dibaryon in Section \ref{SSone}. Lattice results have already begun to accumulate for this system. We can get ahead of the lattice results by beginning to investigate the possibility of bound states with many more quarks.

We have investigated the possibility of additional $I=0$ states constructed from two $\Lambda$ wave functions. Such states can be thought of as multi-combinations of {\it H}-dibaryons with $12, 18, 24\dots$ quarks. Such states would be very difficult if not impossible for the lattice to simulate. Here, it is a simple matter of changing the initial simulation parameters to accommodate these new states. In this model, we formally have a maximum of 4 light quark flavors. We have investigated the 12 quark configuration,
\begin{eqnarray}
N_1=2, g_1=4; N_s=2, g_s=2,\nonumber 
\end{eqnarray}
the 18 quark configuration,
\begin{eqnarray}
N_1=3, g_1=4; N_s=3, g_s=2,\nonumber 
\end{eqnarray}
and the 24 quark configuration,
\begin{eqnarray}
N_1=4, g_1=4; N_s=4, g_s=2.\nonumber
\end{eqnarray}
Higher quark number states are obvious generalizations. We find the following energies and masses in our investigation:\\

\noindent $\bullet$ For the 12 quark case, we find the mass as 4690 MeV/c$^2$, 234 MeV/c$^2$, more than the 4 $\Lambda$ threshold and 62 MeV/c$^2$ more than twice the TF {\it H}-dibaryon mass. \\

\noindent  $\bullet$ For the 18 quark case, we find the mass as 7100 MeV/c$^2$, 416 MeV/c$^2$ more than the 6 $\Lambda$ threshold and 158 MeV/c$^2$ more than three times the TF {\it H}-dibaryon mass.\\

\noindent  $\bullet$ Finally, for the 24 quark case, we find the mass as 9510 MeV/c$^2$, 598 MeV/c$^2$ more than the 8 $\Lambda$ threshold and 254 MeV/c$^2$ more than four times the TF {\it H}-dibaryon mass.\\

These states are moving further away from the bound particle thresholds. Interestingly, the systems are remaining relatively non-relativistic, with a smaller percentage of the component energy being kinetic as more quarks are added. In addition, the spin energies are becoming a smaller system component as well. These results are listed in Table \ref{multiquark}. Again, note that the radius quoted here is the bag radius of the system rather than the charge radius.

\begin{table}
\caption{Multi-quark configuration energies.}

\begin{center} 
\begin{tabular}{|c|c|c|c|c|c|c|c|c|c|}  \hline\hline
Particle            &Bag Radius             &Rest Mass              &{\it  T}            &{\it U}                      &Spin                     &Vol. Energy                 &Total       &Threshold           \\ 
       & $(fm)$         & $(MeV/c^2)$    & $(MeV/c^2)$     &$(MeV/c^2)$          &$(MeV/c^2)$        & $(MeV/c^2)$       &$(MeV/c^2)$           &$(MeV/c^2)$   \\ \hline
\vspace{-.3cm}           &             &             &            &                   &                   &                &      &           \\
$H-{\rm dibaryon}$   &1.58 (1.45) &2174 &386.4 &-280.4 &-61.7 &96.2 &2314 &2228\\ 
\vspace{-.3cm}           &             &             &            &                   &                   &                &      &           \\ \hline
\vspace{-.3cm}           &             &             &            &                   &                   &                &      &           \\
$12-{\rm quark}$       &1.88           &4348 &640.8 &-458.8 &-22.8 &183.2 &4690 &4456\\ 
\vspace{-.3cm}           &             &             &            &                   &                   &                &      &           \\ \hline
\vspace{-.3cm}           &             &             &            &                   &                   &                &      &           \\
$18-{\rm quark}$        &2.18          &6522 &932.5 &-627.3 &-13.9 &286.3 &7100 &6684\\ 
\vspace{-.3cm}           &             &             &            &                   &                   &                &      &           \\ \hline
\vspace{-.3cm}           &             &             &            &                   &                   &                &      &           \\
$24-{\rm quark}$        &2.41         &8696 &1223 &-789.0 &-10.0 &390.0 &9510 &8912\\ 
           &             &             &            &                   &                   &                &      &           \\ \hline\hline
\end{tabular}
\end{center}
\label{multiquark}
\end{table}

\subsection{Nucleon-Nucleon Resonances}\label{SStwo}
When bound states are not present in a channel, our model is useful in predicting multi-quark baryon resonances. In Table \ref{table4} we present a \lq\lq scan" of the configuration energies for the light, two wave function case for nucleon-nucleon scattering. Note these are the configuration energies, not the particle energies. To obtain particle resonance energies, we must proceed by projecting the quantum state into the particle configurations. Linear combinations of these configuration energies will be the actual resonance energies. 

It turns out there are 8 different configurations that we can reach with our two inequivalent wave functions given that $g_1 N_1+g_2 N_2=6$. There are actually 4 light flavors available here,  $u^{\uparrow},u^{\downarrow},d^{\uparrow},d^{\downarrow}$, so that the maximum value of $g_1+g_2$ is 4. Some missing cases are equivalent to those listed. For example, the $(N_1=3, g_1=1; N_2=3,g_2=1)$ case is obviously equivalent to the $(N_1=3, g_1=2)$ case. The explicit possibilities are listed in Table \ref{table4} from highest mass ($N_1=6,g_1=1$) to lowest mass ($N_1=2,g_1=3$). The results are largely as one would expect from the exclusion principle in that states with a greater $N_1$ or $N_2$ are sequentially heavier. We see that all of the states are greater than the two nucleon threshold and that there are no elemental six quark nucleon-nucleon bound states in our model. These could be contributing to the known resonances in both $pp$ and $np$ elastic scattering above the two nucleon threshold. Of course it is impossible to tell if the resonances are from mesons and baryons or true 6 quark states.

\begin{table}
\caption{The eight six-quark configuration energies containing light quarks in the TF quark model. Note that $g_1N_1+G_2N_2=6$.}

\begin{center} 
\begin{tabular}{|c|c|c|c|c|}  \hline\hline
\vspace{-.2cm}&          &        &      &         \\
$N_{1}$            &$g_{1}$             &$N_{2}$              & $g_{2}$            &$Mass$     \\ 
           &             &              &          &$(MeV/c^{2})$     \\  \hline
6   &1 &0 &0 &2309  \\ \hline
5 &1 &1 &1 &2170  \\  \hline
4 &1 &2 &1 &2154 \\  \hline
4 &1 &1 &2 &2125 \\  \hline
3 &2 &0 &0 &2104  \\ \hline
3 &1 &1 &3 &2068 \\  \hline
2 &2 &1 &2 &2005  \\ \hline
2 &3 &0 &0 &2003  \\ \hline
\end{tabular}
\end{center}
\label{table4}
\end{table}

However, there are two surprising aspects to our results. First, the lowest mass state is not the one we expect. One would expect the $(N_1=2, g_1=2; N_2=1,g_2=2)$ state to be lower in mass than the $(N_1=2, g_1=3)$ from the exclusion principle since we are spreading 6 quarks across the maximum number of light flavors. Notice the mass difference here is quite small, $~2$ MeV/c$^2$. Second, the rest energy for the lightest six-quark state is only 89 MeV/c$^2$ above the two-nucleon threshold, using the calculated nucleon mass, 957 MeV/c$^2$, from Table \ref{table3}. This is comparable to the threshold energy found for the {\it H}-dibaryon, which most investigators would expect to be closer to the bound threshold. Note that we only calculate the bare mass here in this quick scan of configuration energies.

\section{Conclusions}\label{seven}

We have presented the first numerical results in the application of the TF model to strong interaction phenomenology. In this paper we have begun to develop the TF quark model as a relatively uncomplicated tool in strong interaction dynamics. Our results are presently restricted to non-relativistic aspects and to two inequivalent TF wave functions. We have shown how to introduce spin into the formalism and found parameters which effectively fit the low energy baryon states. We have begun to understand the internal structure required by the differential equations. In \cite{wilcox} we saw discontinuities at the edge of TF wave functions  due to external pressure. We also now see internal flavor discontinuities, even for degenerate masses, simply due to energy minimization. Such a spatial separation does not occur in traditional non-relativistic models and is a consequence of the statistical nature of the model. 

The results presented here show how to construct the necessary wave functions in the two wave function multi-flavor case with configuration quantum numbers $N_1,g_1,N_2,g_2$. The parameterization found with 5 fit parameters ($\alpha_s, B, m_u =m_d, m_{str}, {\rm g}$) is relatively good, but the isospin violations are too large for baryon multiplets. The squared charge radii of the particles found with our non-relativistic approach are in general too large, but the systematics resulting in a good value of the neutron squared charge radius are encouraging.

We have undertaken three case studies using the low energy phenomenological fits: the {\it H}-dibaryon, nucleon-nucleon resonances and high multi-quark strange states. Our results for strange quark states show growing systems which are remaining non-relativistic and whose spin interactions are a decreasing percentage of the total energy. There are no hints that the systems can be bound. Indeed, the difference between the mass and threshold values in Table \ref{multiquark} are growing as a percentage of the threshold. Strange quark matter scenarios look very unlikely from this point of view. We have also confirmed a phenomenology in nucleon-nucleon scattering consistent with known low energy hadronic physics.  One surprising finding is that the lightest 6 quark light sector mass actually has a threshold value comparable to the equivalent strange sector involving the $\Lambda$. Of course, these findings should be regarded as initial indications rather than robust results. However, in these applications we are consistently finding that the semi-classical statistical point of view does not encourage the idea of bound many-quark baryonic states. 

The phenomenological fits and findings can be viewed in different ways. On one hand the fact that a many-particle theory can accommodate states of three particles is extremely non-trivial. On the other hand one has to keep in mind that one is required to do the fitting in the exact wrong place for such a model. This is similar to situation the $1/N_c$ expansion in strong interaction field theory when $N_c=3$. Keeping this in mind, we are emphasizing the emerging {\it systematics} in the fits rather than the numerical results. Hopefully, lattice calculations can be extended to investigate this picture more completely.

We are just at the beginning point in the application of this model to known strong interaction phenomenology. There are many remaining areas in which to continue the development of the TF quark model. Besides removing the technical restriction to two inequivalent TF flavor wave functions here, we note that the model may be extended:\\

\noindent$\bullet$ To be relativistic \\
\noindent$\bullet$ To include anti-quarks\\
\noindent$\bullet$ To include central heavy quarks\\
\noindent$\bullet$ To examine exotic forms of matter\\

\noindent
All these matters can be addressed within this model. The relativistic equation to be solved is given in Ref.\cite{wilcox}, although the system energy formulas are not yet developed. We can not strongly defend our non-relativistic assumption other than stating that the present model should be regarded as a work in progress. The overly simplistic non-relativistic wave functions used in the spin projection process are leading to unrealistically large isospin splittings within multiplets. This is surely a signal that higher component wave functions should be developed. This would most naturally be done in a relativistic context. The inclusion of anti-quarks is straightforward and would simply be accomplished by introducing a different color coupling for such particles and averaging over the color couplings in the model. Such an extension would allow us to investigate exotic states such as a penta quark $(udud\bar{s})$ state\cite{LEPS} or a four quark $(\bar{c}c\bar{d}u)$ state\cite{BELLE}.  Heavy central quarks can be examined using Coulombic sources much like atomic TF systems. Perhaps most interestingly, the color-flavor locking scenario, involving quark Cooper pairs and massive gluons, can also be explored\cite{cflock}. We believe that the TF quark model can be an effective theoretical tool in delineating the systematics of interesting many-quark baryonic states.

\end{document}